\documentclass[baaa]{baaa}

\usepackage[pdftex]{hyperref}
\usepackage{subfigure}
\usepackage{natbib}
\usepackage{helvet,soul}
\usepackage[font=small]{caption}

\contriblanguage{0}

\contribtype{1}

\thematicarea{11}

\title{Nuevas estrategias de enseñanza: unidades didácticas basadas en temas de la Astronomía Cultural}

\titlerunning{Nuevas estrategias de enseñanza: unidades didácticas basadas en temas de la Astronomía Cultural}

\author{
F. A. Karaseur\inst{1,2}, J. I. Bastero\inst{1,2}, S. J. Gar\'ofalo\inst{2}
\&
A. Gangui\inst{2,3}
}

\authorrunning{Karaseur, Bastero, Garófalo \& Gangui}

\contact{fkaraseur@hotmail.com}

\institute{
Universidad Nacional del Centro de la Provincia de Buenos Aires, Argentina \and
Instituto de Formaci\'on e Investigaci\'on en Ense\~nanza de las Ciencias, Facultad de Ciencias Exactas y Naturales, UBA, Argentina \and
Instituto de Astronom\'ia y F\'isica del Espacio, CONICET--UBA, Argentina
}

\resumen{Una de las principales dificultades que presentan los estudiantes en el aprendizaje de tópicos de astronomía es que no logran relacionar la información teórica con lo que experimentan en el mundo que los rodea. La construcción por parte de los estudiantes de un marco conceptual acorde al modelo científico astronómico demanda cambios de enfoque en la enseñanza actual. En este marco, la Astronomía Cultural (AC) es una disciplina de la cual podemos valernos para repensar nuevas estrategias didácticas. En este trabajo se presentan dos propuestas contextualizadas desde la AC. En la primera, se aborda la enseñanza de conceptos de espacio y tiempo mediante ejemplos tradicionales de orientación por las estrellas y del empleo del calendario, utilizando el caso de la navegación oceánica histórica sin instrumentos avanzados ya destacado en estudios etnoastronómicos. En la segunda, estos conceptos son trabajados a partir de un estudio de caso, esta vez arqueoastronómico: el calendario de horizonte monumental del sitio arqueológico Chanquillo, para luego continuar con la identificación local de marcadores de horizonte que permitan a los estudiantes la construcción de sus propios calendarios. Se pretende así ilustrar modos de introducir elementos de la AC en unidades didácticas que tengan como uno de sus principales objetivos que los estudiantes logren establecer correspondencias entre construcciones propias del micro al megaespacio que los rodea.}

\abstract{One of the main difficulties that students have in learning astronomy topics is that they fail to relate theoretical information with what they experience in the world around them. The construction by students of a conceptual framework in accordance with the astronomical scientific model demands changes in the current teaching approach. Within this framework, Cultural Astronomy (CA) is a discipline that we can use to rethink new didactic strategies. This paper presents two contextualized proposals from CA. In the first one, the teaching of space and time concepts is approached through traditional examples of orientation by the stars and the use of the calendar, using the case of historical ocean navigation without advanced instruments already highlighted in ethnoastronomical studies. In the second, these concepts are worked on from a case study, this time archaeoastronomical: the monumental horizon calendar of the Chankillo archaeological site, and then continue with the local identification of horizon markers that allow students to build their own calendars. The aim is to illustrate ways of introducing CA elements in didactic units that have as one of their main objectives that the students manage to establish correspondences between constructions of the micro to the mega-space that surrounds them.}


\begin{document}

\maketitle

\section{Introducción}\label{S_intro}

Los diseños curriculares de los niveles primario, secundario y superior de la Ciudad Autónoma de Buenos Aires prescriben la enseñanza de conceptos de espacio y tiempo vinculado con los movimientos del Sol, otras estrellas y demás astros. Aun así, estudios de investigación en la enseñanza y aprendizaje de la astronomía muestran las dificultades que tienen los estudiantes para relacionar la información teórica con lo que experimentan en el mundo que los rodea. Las mismas pueden atribuirse a una planificación de la enseñanza sustentada  en una metodología tradicional, que prescinde de la consideración de los conocimientos previos de los estudiantes para la construcción de los modelos de ciencia escolar correspondientes \citep{Porlan_1999}. Desde esta última perspectiva, buscamos establecer correspondencias entre la observación directa del cielo, la construcción del pensamiento espacial y, simultáneamente, el desarrollo de capacidades como la descripción, la definición y la justificación \citep{GonzalezRodriguez_etal_2015},\citep{Plummer_2014}.

Nuestras propuestas parten de poner en conflicto las concepciones que prevalecen en los estudiantes. Entre ellas, no reconocer las variaciones en las trayectorias que describen los astros, ni el hecho de que las mismas son propias del punto de vista espacio-temporal del observador. En el caso particular del Sol, esto se traduce en concepciones de acimuts de salida y puesta fijos y culminaciones superiores que alcanzan el cénit, independientemente de la posición del observador sobre la Tierra y el día del año \citep{Gangui_Iglesias_2015}. 

La planificación de la enseñanza que se presenta en este trabajo promueve, en línea con trabajos anteriores \citep{Karaseur_Gangui_2021}, que los estudiantes establezcan para estos fenómenos correspondencias entre posibles representaciones y su aprehensión a gran escala, es decir, construyan modelos de estos fenómenos del micro al megaespacio que los rodea \citep{Lanciano_2014}.    
Dentro de los propósitos actuales de la enseñanza de las ciencias naturales se encuentra el de enfatizar el conocimiento científico como una construcción social-histórica, que permite comprender y operar sobre el entorno. En este marco resulta propicia la enseñanza de las ciencias naturales en contexto, dado que esto favorece el interés por aprender, así como también la adquisición de competencias científicas en torno al pensamiento crítico \citep{Caamanno_2011},\citep{DeJong_2015}. Es en  este sentido que la incorporación de la Astronomía Cultural a las planificaciones de enseñanza cobra especial relevancia al considerar que es un área que busca entender las distintas formas en las que los objetos y fenómenos del cielo se registran, influyen, impactan y guían las creencias, los sistemas de conocimiento y las tradiciones culturales \citep{Lopez_Hamacher_2017}. Dicha área de conocimiento permite poner de manifiesto que las ciencias están sometidas a ciertos condicionantes y determinadas por la sociedad en la que se desarrollan y, además, ayuda a valorar y a entender la importancia de las relaciones de la astronomía con la sociedad y la tecnología \citep{Palomar_Solbes_2015},\citep{Jafelice_2015}.  

El presente trabajo pretende ilustrar los aportes que la astronomía cultural podría ofrecer a la enseñanza de la astronomía, en este caso incorporando dos recursos como puertas de entrada
aunque sin profundizar explícitamente en aspectos culturales e históricos como sería deseable \citep{Rodrigues_Leite_2020}. Por un lado, el uso de narrativas oceánicas que destacan
algunos de los vínculos entre la astronomía y la navegación no instrumental, principalmente la orientación por las estrellas \citep{Pimenta_2015}. Por otro lado, las evidencias históricas
de las Torres de Chanquillo como marcadores de los puntos de salida y puesta del Sol a lo largo del año desde un punto de vista en particular, es decir, como calendario de horizonte
\citep{Gangui_2015}.

Aunque estas propuestas de enseñanza fueron pensadas para ser implementadas principalmente en Ciencias Naturales de secundaria, las dos pueden ser adaptadas también para ser llevadas al aula en institutos de formación docente. Para ambas propuestas los alumnos requieren como contenidos previos conocer los puntos cardinales y el movimiento diario del Sol visto por un observador local, al igual que conocimientos mínimos sobre el sistema de coordenadas horizontal, las constelaciones más usuales y los momentos característicos del ciclo solar, como los equinoccios.

\section{Propuesta de enseñanza: narrativas oceánicas y el movimiento de las estrellas}

\subsection{Primer momento didáctico}
El objetivo de esta instancia es que los estudiantes expliciten sus ideas previas a partir del planteo de situaciones problematizadoras que requieran utilizar a las estrellas como referencia para la ubicación geográfica y temporal. Para ello, se presenta como puerta de entrada un texto adaptado de “Vigilia del Almirante”, de Augusto Roa Bastos: 

“El círculo luminoso del mediodía transforma el Sol en oro cenital. Su nadir, la miseria y la muerte (…) El Sol está en Libra. Hubiera preferido que estuviera en Gémino. Estamos atravesando los últimos fuegos del equinoccio (…) A los navegantes nos están reservados los fríos…al otro lado del mundo. ¿No es la mejor prueba de que la Tierra…es redonda? (…) En este punto del hemisferio, la Polar no deja ver ya su luz astral. Otras constelaciones la han reemplazado.” 

Luego de su lectura, pueden formularse preguntas como: “¿Será posible determinar el lugar y el momento del año donde sucede esta historia?”.  

\subsection{Segundo momento didáctico}
Este momento tiene como finalidad que los estudiantes expliciten y avancen en la comprensión del movimiento del Sol a lo largo del año y en distintas latitudes. Se espera que sean capaces de determinar la fecha y el lugar aproximados donde transcurre la historia. Para ello, se propondrá a los estudiantes que expliciten sus ideas representando el movimiento del Sol que ven a lo largo de un día y cómo se lo imaginan seis meses más tarde. Luego contrastarán con observaciones sistemáticas directas, con Mapa Estelar y con Stellarium. Finalmente, el docente podría invitarlos a proponer posibles momentos del año en el que transcurre el relato, recurriendo a materiales concretos como linterna, esferas de telgopor, palitos de brochette y fósforos, además de tablas de efemérides y Solar System Scope. Con la misma dinámica se propone que infieran posibles lugares. Para ello se podría utilizar algún análogo concreto como el globo terráqueo paralelo. De esta forma el docente estimularía el análisis de la relación entre la forma de la Tierra y las estaciones, modelizando situaciones de navegación sobre una Tierra plana o bien sin inclinación de su eje con respecto de la eclíptica. 

\subsection{Tercer momento didáctico}
Su objetivo es promover la reflexión metacognitiva y la transferencia de conocimiento a otras situaciones similares. Para ello, se formularán preguntas como: “¿Qué logré aprender a partir del relato? ¿Qué dificultades tuve?”. Como evaluación formativa se propone que los estudiantes reelaboren el relato situándose imaginariamente no en alta mar sino en su propia ciudad, por ejemplo, durante el mes de marzo, cuando el Sol se halla en la constelación diametralmente opuesta a la de Libra (que señala el texto de Roa Bastos), apoyándose en registros de observaciones propias como esquemas o fotos (Fig. 1).  

\begin{figure}[!t]
\centering
\includegraphics[width=\columnwidth]{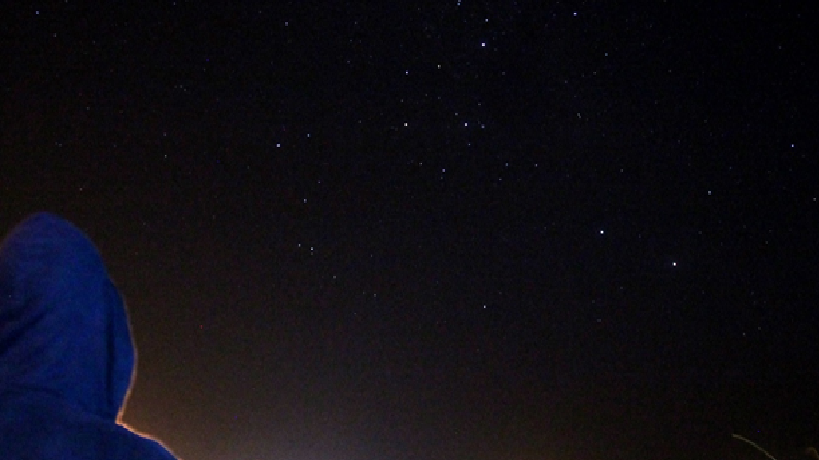}
\caption{La “Cruz del Sur”, constelación característica del hemisferio sur (fotografía de los autores).}
\label{Figura1}
\end{figure}

\section{Propuesta de enseñanza: las torres de Chanquillo y calendarios de horizonte}

\subsection{Primer momento didáctico}
Este primer momento está destinado a conocer qué ideas tienen los estudiantes acerca del arco solar diurno y su relación con las estaciones y la latitud con el fin de utilizarlas como punto de partida. Para ello, se propone utilizar como puerta de entrada la imagen de la Fig. 2. acompañada de preguntas como: “¿Será posible determinar en qué momento del año se realizó esta fotografía? ¿Cambiará algo si el observador se orienta de otro modo respecto de las torres?”. 
 
\subsection{Segundo momento didáctico}
Este momento tiene como finalidad que los estudiantes reconozcan los cambios en los acimuts de salida y puesta del Sol a lo largo del año en las torres de Chanquillo y trabajar en el salto cognitivo necesario que les permita luego transferir el conocimiento logrado a su localidad geográfica. Para ello, el docente propondrá una actividad en grupos en la que a partir de fotografías deberán identificar momentos del día y vincularlos con el punto de observación. Luego se pedirá que anticipen el momento en el que el Sol pasará sobre otra torre diferente de aquella que se muestra en la Fig. 2. Contrastarán sus hipótesis con Stellarium y publicaciones de arqueoastronomía. Discutirán diferencias y similitudes acerca de lo que verían si estas torres estuvieran localizadas en su propia ciudad. 
\subsection{Tercer momento didáctico}
Se busca promover la reflexión metacognitiva y la transferencia de conocimiento a nuevas situaciones. Para ello, se formularán preguntas como: “¿Qué logré aprender de las torres? ¿Cómo influyó el trabajo en grupos para la resolución de las actividades?” Como evaluación formativa los estudiantes tomarán fotografías de las posiciones del Sol en horarios cercanos a la salida o la puesta desde un punto fijo de observación y orientación, para realizar inferencias sobre las fotos de sus compañeros.

\begin{figure}[!t]
\centering
\includegraphics[width=\columnwidth]{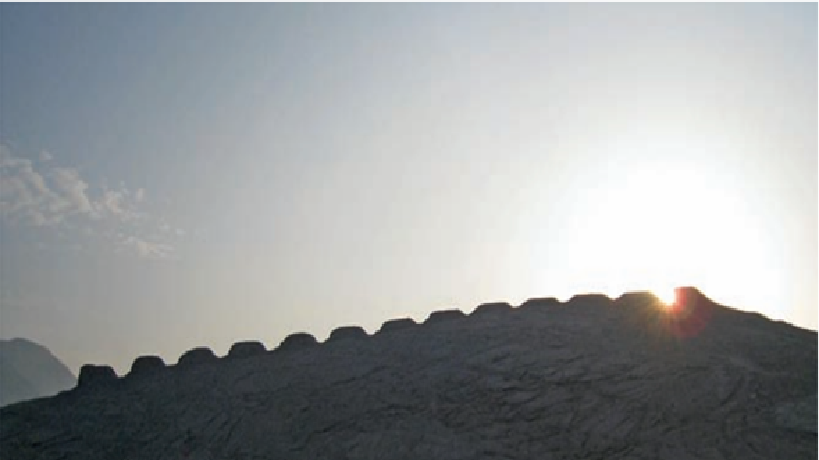}
\caption{Las trece torres de Chanquillo, ubicadas cerca de la costa de Perú, fueron utilizadas en la época preincaica como calendario de horizonte (fotografía de los autores).}
\label{Figura2}
\end{figure}

\section{Conclusión}

En este trabajo se presentan dos propuestas de enseñanza progresiva de temas astronómicos contextualizándolos con elementos de la astronomía cultural. Este enfoque histórico-epistemológico, que aún resta profundizar, acerca a los estudiantes a conocer no sólo cómo la humanidad fue construyendo los modelos científicos actuales, sino también cómo dicho conocimiento es siempre un conocimiento situado, ligado a quienes somos \citep{Lopez_Hamacher_2017}. En futuros trabajos esperamos presentar los resultados de la implementación de estas propuestas de enseñanza en contextos reales con el objetivo de evaluar los aprendizajes logrados.      


\bibliographystyle{baaa}
\small
\bibliography{717_v3}

\begin{thebibliography}{14}
\providecommand{\natexlab}[1]{#1}

\bibitem[{{Caamaño}(2011)}]{Caamanno_2011}
{Caamaño} A., 2011, Alambique, 69, 21

\bibitem[{{De Jong}(2015)}]{DeJong_2015}
{De Jong} O., 2015, Educació Química EduQ, 20, 25

\bibitem[{{Gangui}(2015)}]{Gangui_2015}
{Gangui} A., 2015, Ciencia Hoy, 25, 55

\bibitem[{{Gangui} \& {Iglesias}(2015)}]{Gangui_Iglesias_2015}
{Gangui} A., {Iglesias} M.C., 2015, \textit{{Didáctica de la astronomía:
  actualización disciplinar en ciencias naturales. Propuestas para el aula}},
  Paid\'os, Ciudad Autónoma de Buenos Aires

\bibitem[{{Gonzalez Rodriguez} et~al.(2015){Gonzalez Rodriguez}, {Garcia
  Barros} \& {Martinez}}]{GonzalezRodriguez_etal_2015}
{Gonzalez Rodriguez} C., {Garcia Barros} S., {Martinez} C., 2015, Enseñanza de
  las Ciencias, 33, 71

\bibitem[{{Jafelice}(2015)}]{Jafelice_2015}
{Jafelice} L., 2015, Revista Latino-Americana de Educação em Astronomia, 19,
  57

\bibitem[{{Karaseur} \& {Gangui}(2021)}]{Karaseur_Gangui_2021}
{Karaseur} F.A., {Gangui} A., 2021, Proceedings of IAU Symposium S367, 15, 444

\bibitem[{{Lanciano}(2014)}]{Lanciano_2014}
{Lanciano} N., 2014, \textit{{Ensino de Astronomia na escola: concepções,
  ideias e práticas}}, 169--195, Átomo, Uberlândia

\bibitem[{{Lopez} \& {Hamacher}(2017)}]{Lopez_Hamacher_2017}
{Lopez} A.M., {Hamacher} D., 2017, Revista Ciencia Y Tecnolog\'ia, 19, 11

\bibitem[{{Palomar} \& {Solbes}(2015)}]{Palomar_Solbes_2015}
{Palomar} R., {Solbes} J., 2015, Enseñanza de las Ciencias, 33, 91

\bibitem[{{Pimenta}(2015)}]{Pimenta_2015}
{Pimenta} F., 2015, \textit{{Handbook of Archaeoastronomy and Ethnoastronomy}},
  43--65, Springer-Verlag, New York

\bibitem[{{Plummer}(2014)}]{Plummer_2014}
{Plummer} J., 2014, Studies in Science Education, 50, 1

\bibitem[{{Porlan}(1999)}]{Porlan_1999}
{Porlan} R., 1999, \textit{{Enseñar Ciencias Naturales. Reflexiones y
  propuestas didácticas}}, 23--64, Paidós, Ciudad Autónoma de Buenos Aires

\bibitem[{{Rodrigues} \& {Leite}(2020)}]{Rodrigues_Leite_2020}
{Rodrigues} M., {Leite} C., 2020, Ensaio, 22, 1

\end{thebibliography}
 
\end{document}